\documentclass[acmlarge,screen]{acmart}
\usepackage{multirow}
\AtBeginDocument{%
  \providecommand\BibTeX{{%
    \normalfont B\kern-0.5em{\scshape i\kern-0.25em b}\kern-0.8em\TeX}}}

\setcopyright{acmcopyright}
\copyrightyear{2024}
\acmYear{2024}
\acmDOI{XXXXXXX.XXXXXXX}

\acmConference[IMWUT '24]{Proceedings of the ACM on Interactive, Mobile, Wearable and Ubiquitous Technologies}{October 5--9 2024}{Melbourne, Australia}
\acmBooktitle{IMWUT '24: Proceedings of the ACM on Interactive, Mobile, Wearable and Ubiquitous Technologies, October 5--9 2024, Melbourne, Australia} 
\acmPrice{15.00}
\acmISBN{978-1-4503-XXXX-X/18/06}

\begin{document}

\title[AutoTask]{AutoTask: Executing Arbitrary Voice Commands by Exploring and Learning from Mobile GUI}
% AutoTask: Executing Unknown Voice Command by Exploring and Learning from Mobile GUI
% AutoTask: a Mobile Voice Command Interface for Executing Out-of-Vocabulary Intentions

\author[1]{Lihang Pan}
\authornotemark[1]
\orcid{0000-0001-8856-0309}
\email{plh18@mails.tsinghua.edu.cn}
\author[1]{Bowen Wang}
\authornote{Equal contribution}
\email{wbw20@mails.tsinghua.edu.cn}
\orcid{0009-0009-1358-722X}
\affiliation{%
  \institution{Department of Computer Science and Technology, Tsinghua University}
  \city{Beijing}
  \country{China}
}
\author{Chun Yu}
\orcid{0000-0003-2591-7993}
\authornote{Corresponding author.}
\affiliation{%
  \institution{Department of Computer Science and Technology, Tsinghua University}
  \city{Beijing}
  \country{China}
}

\author{Yuxuan Chen}
\affiliation{%
  \institution{Department of Computer Science and Technology, Tsinghua University}
  \city{Beijing}
  \country{China}
}

\author{Xiangyu Zhang}
\orcid{0009-0006-2183-3054}
\affiliation{%
  \institution{Department of Computer Science and Technology, Tsinghua University}
  \city{Beijing}
  \country{China}
}

\author{Yuanchun Shi}
\orcid{0000-0003-2273-6927}
\affiliation{%
  \institution{Department of Computer Science and Technology, Tsinghua University}
  \city{Beijing}
  \country{China}
}
\email{shiyc@tsinghua.edu.cn}

% \renewcommand{\shortauthors}{Anonymous}

\begin{abstract}
Voice command interfaces (VCIs) have gained increasing importance, enabling hands-free and eyes-free interaction with digital devices. However, the inherent complexity in constructing effective voice interfaces has limited the VCIs' functionalities to only a small fraction of GUI applications and tasks. This paper presents AutoTask, a VCI capable of automating any task in any mobile application without configuration or modification from developers or end users. The primary challenge for AutoTask is the lack of knowledge, as it needs to accomplish unknown tasks (e.g., user commands) within an unknown environment (e.g., GUI). To address this challenge, AutoTask employs two strategies: (1) trial and error: AutoTask explores the GUI, attempts potential operation sequences, and recovers from errors through backtracking; (2) learning from the environment: AutoTask accumulates experiences during exploration and summarizes correct knowledge from these experiences. We implemented AutoTask on Android devices and conducted an evaluation study, which proved the feasibility of AutoTask.
\end{abstract}




\begin{teaserfigure}
  \includegraphics[width=\textwidth]{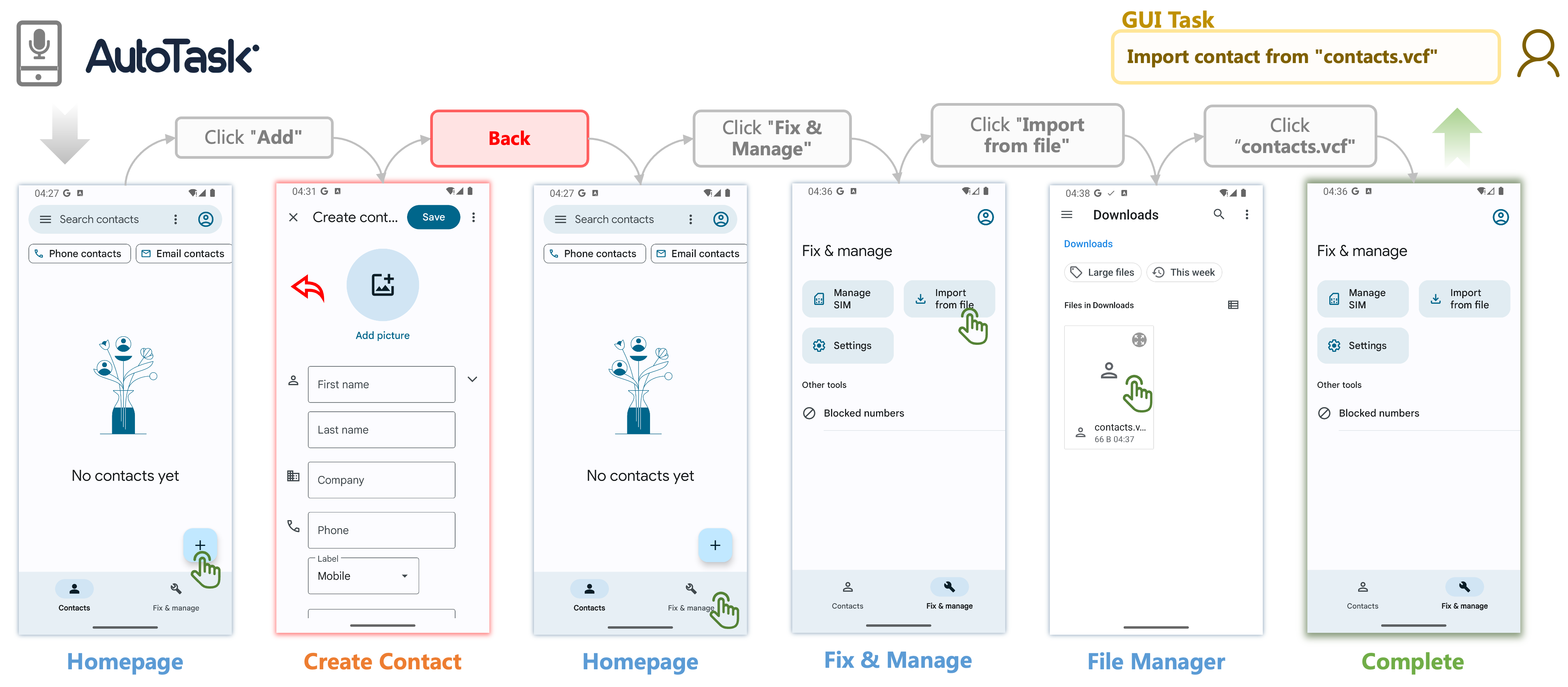}
  \caption{To execute the command "import contacts from contacts.vcf" in the Contacts application (version 4.8.17), AutoTask first clicks the "Add" button, transitioning from Page 1 to Page 2. AutoTask finds that it can only manually add a contact on Page 2; hence, it reverts to Page 1 and clicks another button labeled "Fix \& Manager". It then completes the import process through subsequent steps (4 \& 5). After finishing the task, AutoTask synthesizes knowledge from its experiences, improving its ability for future commands.}
  \label{fig: teaser}
\end{teaserfigure}

\settopmatter{printacmref=false}
\setcopyright{none}
\renewcommand\footnotetextcopyrightpermission[1]{}
\pagestyle{plain}

\maketitle

\section{Introduction}
Voice interaction can effectively enhance the interactivity of applications, enabling end users to automate multi-step GUI tasks (e.g., setting an alarm for 9 AM) eyes-freely and hands-freely \cite{pan2022automatically, wulf2014hands, zhong2014justspeak, luger2016like, cowan2017can}. However, constructing voice command interfaces (VCIs) for existing GUI tasks is challenging and requires significant effort \cite{pan2022automatically, 10.1145/3610929}. Consequently, existing VCIs support only a limited set of predefined intents, failing to cover the actual needs of users \cite{luger2016like, pan2022automatically}.

Large language models (LLMs) have been applied to numerous domains and have significantly reduced the cost of system development and deployment \cite{ross2023programmer, zheng2023towards, fan2023large, fiannaca2023programming, chung2022talebrush}. While LLMs help understand user commands and alleviate the development burden of VCIs, their application is limited in scope \cite{wang2023enabling}. Developers are still required to pre-define a set of intents for the voice interface \cite{arsan2021app, pan2022automatically}. Additionally, they must configure how these intents are executed \cite{li2018kite, li2017sugilite} and address potential errors arising from LLMs \cite{tandon-etal-2022-learning}.

In this paper, we present AutoTask, a ready-to-use VCI that operates without any modifications or configurations by either developers or end users. It is capable of executing any intent within any application. AutoTask accomplishes arbitrary tasks (i.e., user commands) in an unknown environment (i.e., the GUI), the primary challenge of which is the lack of necessary knowledge. To overcome this, AutoTask (1) engages in trial and error: exploring the GUI, attempting possible operation sequences, and recovering from errors through backtracking; and (2) learns from the environment by accumulating experiences during exploration and summarizing knowledge from them.

As illustrated in Figure \ref{fig: teaser}, AutoTask comprehends the semantics of the GUI and the user command and determines an operations sequence to carry out the given task. For example, to execute the command "Import contacts from contacts.vcf", AutoTask first chooses to click the "Add" button on Page 1. It then emulates this operation on the GUI, leading to content updates and a transition from Page 1 to Page 2. AutoTask subsequently evaluates the correctness of the executed operation sequence. If an error is detected, AutoTask revokes its actions to rectify it. For instance, upon reaching Page 2, AutoTask recognizes that it can only manually add a single contact there and cannot perform a batch import from a file. Consequently, it reverts to Page 1 and selects another button labeled "Fix \& Manager". This process continues until the task is successfully completed. Additionally, AutoTask improves its performance by accumulating experiences while navigating the GUI and summarizing knowledge from these experiences, including:
\begin{enumerate}
    \item Environmental knowledge: for example, in Figure \ref{fig: teaser}, AutoTask can learn that clicking the "Add" button does not lead to batch importing of contacts. This knowledge can expedite AutoTask's execution of subsequent commands.
    \item Task knowledge: for instance, in Figure \ref{fig: teaser}, AutoTask can learn that the intent "Import contacts from a file" requires a parameter specifying the file name. This knowledge aids AutoTask in understanding the semantics of the commands.
    \item Execution knowledge: as shown in Figure \ref{fig: teaser}, AutoTask can learn the correct operation sequence for importing contacts from a file. This sequence can be directly replayed to accomplish similar tasks and help execute other commands.
\end{enumerate}

This paper makes two main contributions:
\begin{enumerate}
    \item We introduce a new paradigm in which an agent accomplishes unknown tasks in an unknown environment. The agent explores the environment to find a solution and summarizes its experiences into knowledge to enhance its capabilities.
    \item We present a ready-to-use VCI named AutoTask, where end users can automate any intent with a single command. Experimental results proved its usability. We implemented AutoTask on Android smartphones and conducted an evaluation study. The experimental results proved its usability.
\end{enumerate}
\section{Related Work}
Voice command interfaces can effectively reduce the interaction burden, enabling users to interact with devices hands-freely and eyes-freely \cite{pan2022automatically, wulf2014hands, zhong2014justspeak, luger2016like, cowan2017can}. However, constructing a VCI for mobile devices requires significant effort \cite{pan2022automatically, li2018kite, 10.1145/3610929}, leading to the existing VCIs covering only a limited set of GUI functionalities. The workload primarily encompasses determining the supported intent set, understanding user natural language commands accurately, and executing tasks correctly. Self-improvement of VCIs during runtime is a crucial approach to reducing effort \cite{pan2022automatically} but has yet to gain widespread support. Table \ref{table: compare} compares existing VCIs with AutoTask in these four aspects.

\begin{table}[ht]
\caption{Comparison of AutoTask and existing VCIs. The content within parentheses indicates the workload required by developers or end users. PBD stands for programming by demonstration, and CCG stands for combinatory categorial grammar.}
\label{table: compare}
\begin{tabular}{|c|c|c|c|c|}
\hline
 &
  Intent Set &
  Command Understanding &
  Task Execution &
  Self-Improvement \\ \hline
\begin{tabular}[c]{@{}c@{}}SUGILITE\\ \cite{li2017sugilite}\end{tabular} &
  \begin{tabular}[c]{@{}c@{}}Predefined\\ (PBD)\end{tabular} &
  \begin{tabular}[c]{@{}c@{}}CCG \\ (Handcrafted rules)\end{tabular} &
  \begin{tabular}[c]{@{}c@{}}Replay operations\\ (PBD)\end{tabular} &
  \begin{tabular}[c]{@{}c@{}}Improve executing\\ (PBD)\end{tabular} \\ \hline
\begin{tabular}[c]{@{}c@{}}SAVANT\\ \cite{arsan2021app}\end{tabular} &
  \begin{tabular}[c]{@{}c@{}}Predefined\\ (specify a list)\end{tabular} &
  \begin{tabular}[c]{@{}c@{}}Dialogflow agents\\ (provide examples)\end{tabular} &
  \begin{tabular}[c]{@{}c@{}}Search app screen\\ (collect traces)\end{tabular} &
  Not supported \\ \hline
\begin{tabular}[c]{@{}c@{}}AutoVCI\\ \cite{pan2022automatically}\end{tabular} &
  \begin{tabular}[c]{@{}c@{}}Predefined\\ (PBD)\end{tabular} &
  \begin{tabular}[c]{@{}c@{}}BERT\\ (extra dialogues)\end{tabular} &
  \begin{tabular}[c]{@{}c@{}}Replay operations\\ (PBD)\end{tabular} &
  \begin{tabular}[c]{@{}c@{}}Improve understanding\\ (extra dialogues)\end{tabular} \\ \hline
\begin{tabular}[c]{@{}c@{}}LLM4Mobile\\ \cite{wang2023enabling}\end{tabular} &
  \begin{tabular}[c]{@{}c@{}}Intents in the GUI\\ (No effort)\end{tabular} &
  \begin{tabular}[c]{@{}c@{}}LLM\\ (No effort)\end{tabular} &
  N/A &
  Not supported \\ \hline
    AutoTask &
  \begin{tabular}[c]{@{}c@{}}Any intent\\ (No effort)\end{tabular} &
  \begin{tabular}[c]{@{}c@{}}LLM\\ (No effort)\end{tabular} &
  \begin{tabular}[c]{@{}c@{}}Explore the GUI\\ (No effort)\end{tabular} &
  \begin{tabular}[c]{@{}c@{}}Improve understanding\\ Improve executing \\ (No effort)\end{tabular} \\ \hline
\end{tabular}
\end{table}

\subsection{Supported intents of VCIs}
The earliest voice interfaces on smartphones only supported single-step GUI operations \cite{ashok2015capti, ashok2017web, zhong2014justspeak, wang2023enabling, vu2023voicify}, for example, clicking a button already present on the GUI. The intent sets of this kind of VCI are limited to the current GUI contents. Although the VCIs can be applied to any mobile application without any configuration or modification, GUI tasks typically require multiple operations (e.g., clicking several buttons sequentially), and providing voice commands for each step would impose a significant interaction burden. Therefore, this approach is mainly used for accessibility purposes \cite{grussenmeyer2017accessible, zhong2014justspeak, 10.1145/2468356.2468706} and has not been widely adopted by ordinary users.

Task-oriented VCIs \cite{li2017sugilite, arsan2021app, li2018kite, pan2022automatically, sereshkeh2020vasta} (e.g., Siri) address the aforementioned issues and reduce the interaction burden. End users only provide a single voice command, and the virtual assistant can automatically complete a multi-step task on the GUI (e.g., setting a 9:00 am alarm). However, existing task-oriented voice assistants only support a limited set of intents. For instance, Siri does not support sending WhatsApp messages. This results in two challenges: on one hand, developers need to invest a significant amount of effort (e.g., conducting formative studies \cite{li2017sugilite, arsan2021app}) to determine a useful intent set; on the other hand, end users often complain about the discoverability of functionalities \cite{murad2019revolution, srinivasan2019discovering, corbett2016can, yankelovich1996users} and the lack of support for necessary intents \cite{luger2016like}.

AutoTask differs significantly from the two categories of voice interfaces mentioned above. AutoTask is a task-oriented voice assistant capable of automating multi-step tasks with a single command. However, AutoTask does not rely on a predefined set of intents and can accommodate any intent that can be executed on the GUI without requiring additional overhead from developers or end users.

\subsection{Understanding User Commands}
A task-oriented voice assistant understands the user command to: (1) classify the command into a specific intent (e.g., sending a message); and (2) identify parameters for the intent (e.g., message content and message recipient) \cite{pan2022automatically, li2017sugilite, li2018kite}. The most traditional approach to command understanding is using context-free grammar (e.g., regular expressions \cite{ashok2017web, zhong2014justspeak} and combinatory categorial grammar (CCG) \cite{azaria2016instructable, li2017sugilite, li2018appinite, li2019pumice}). Additionally, researchers have employed more sophisticated algorithms (e.g., word dependency \cite{sereshkeh2020vasta}, n-grams \cite{gao2015datatone, kim2019vocal} and word embedding \cite{sereshkeh2020vasta}) to extract features from commands and create natural language processing scripts. These solutions heavily rely on handcrafted rules \cite{ward1994recent, ashok2015capti}, necessitating significant developer effort; however, their performance is relatively poor \cite{azaria2016instructable, luger2016like, xu2020clue, yankelovich1995designing} and cannot satisfy users' needs.

Nowadays, many systems utilize deep neural networks to comprehend user commands, achieving satisfactory results \cite{qu2019bert, seo2019real}. Before the widespread adoption of pre-trained models, researchers needed to collect a large amount of training data \cite{brown2020language, devlin2018bert, thomason2015learning} and meticulously fine-tune the model's architecture and parameters to achieve good results \cite{zhang2015sensitivity}. This process involved a significant workload. With the development of pre-trained models, researchers only provide prompts and a few optional examples, and large language models (LLMs) can successfully understand the commands \cite{wang2023enabling, liu2023wants}.

An often overlooked issue is that interactive systems may inaccurately interpret user commands, leading to conversation breakdowns \cite{beneteau2019communication, grudin2019chatbots, pan2022automatically}. This problem increases the user's burden \cite{jiang2013users, pelikan2016nao} and reduces their willingness to engage in interactions \cite{jain2018evaluating, luger2016like, zamora2017m}. Existing solutions all entail additional costs. For instance, developers can programmatically address breakdowns \cite{bohus2005error, kim2019did}. AutoVCI \cite{pan2022automatically} and SOVITE \cite{li2020multi} require additional user interactions to ensure accurate command understanding.

AutoTask utilizes LLMs to comprehend user commands, enabling support for any task in any application with minimal development effort. To address the issue of LLM errors, AutoTask does not require developer or user involvement; instead, it automatically learns from the mobile GUI and continually adjusts its command understanding results.

\subsection{Executing the Command}
Existing VCIs search for and execute scripts in databases based on the command understanding results; the scripts can automatically carry out the user commands. These scripts can be categorized into two types based on their origins: those created by developers and those generated by end users.

The scripts for the majority of commercial voice assistants are manually created by developers \cite{ravindranath2012code}. For example, Siri directly invokes functions implemented by developers to execute voice commands. Because developers need to write a script for each intent, this approach significantly limits the number of functionalities available in the voice interface \cite{pan2022automatically}.

Since end users have a strong demand for voice assistants that can support their personalized needs, researchers have proposed different methods to collect execution scripts from users. One typical approach is program by demonstration (PBD) \cite{pan2022automatically, allen2007plow, cypher1993watch, sugiura1996simplifying, myers1986visual, li2017sugilite, li2018kite, sereshkeh2020vasta, dey2004cappella, 10.1145/3397329}. Users can demonstrate how they complete tasks on the GUI; this process is recorded and automatically transformed into execution scripts. Researchers have also attempted to extract execution scripts from users' historical behavior records automatically \cite{arsan2021app, antila2012routinemaker, lau2010conversational}. While this approach avoids the explicit burden of demonstration, it also results in unpredictable system capabilities and relies on the time and quality of data accumulation. 

Note that scripts collected from users are not always correct. For example, replaying the operation sequence demonstrated by users may fail due to pop-up windows or application version updates \cite{li2017sugilite}. Existing solutions require users to handle exceptions manually \cite{li2017sugilite}, which introduces additional burdens.

AutoTask does not require any predefined scripts, whether they originate from developers or end users. It dynamically calculates a potential operation sequence on the GUI at runtime. Furthermore, it assesses whether the sequence is correct and automatically handles errors. The entire process does not necessitate any user intervention.

\subsection{Self-Improvement}
Many systems can learn and improve themselves through interactions with users. A typical application scenario is learning user preferences to provide better services (e.g., recommendations \cite{massimo2017learning, tu2019fingerprint}, navigation \cite{rana2020navigation}, and scheduling \cite{gervasio2005active}). AutoVCI \cite{pan2022automatically} can enhance its semantic understanding ability through multi-turn dialogues with users \cite{thomason2015learning}. In the field of machine learning, this approach is known as "human-in-the-loop" \cite{stiennon2020learning, ziegler2019fine, wu2022survey}, in which users contribute to improving machine capabilities by providing annotations. This approach has achieved significant success in training large language models \cite{openai2023gpt4, ouyang2022training}.

Reinforcement learning is a common approach for improving machine capabilities without user intervention: AI-driven agents enhance themselves based on rewards provided by the environment \cite{kaelbling1996reinforcement}. This approach has been widely applied in fields such as gaming \cite{kaiser2019model} and autonomous driving \cite{kiran2021deep}, but it has seen limited application in executing user commands on GUIs \cite{branavan2010reading}. AppBuddy \cite{shvo2021appbuddy} is a preliminary attempt in this direction; however, it suffers from issues like sparse rewards \cite{li2023zeroshot} and excessive trial-and-error steps, making it unsuitable for direct application in interactive systems. AutoTask shares a similar concept with reinforcement learning: it autonomously summarizes knowledge from its explorations of the GUI, all without requiring user intervention.
\section{Problem Formulation \& Solution}
This paper focuses on problems of the following form: an intelligent agent is required to complete an unknown task in an unknown environment. This type of problem is prevalent in applying artificial intelligence (AI) to daily tasks for two primary reasons. Firstly, it is impossible for the developers to collect corpora and pre-train an agent for every real-world scenario. Secondly, end users often struggle to, or choose not to, provide comprehensive, structured descriptions and step-by-step procedures for tasks. Solving problems characterized by these patterns can significantly broaden the application of AI in everyday tasks.

AutoTask is a solution to the problem in the field of VCIs: it automates the execution of interaction intents (i.e., tasks) expressed through natural language commands by simulating user operation sequences within the GUI (i.e., the environment) of mobile devices. AutoTask supports any applications and intents, ensuring comprehensive coverage of GUI tasks for voice assistants.

The core challenge of "completing unknown tasks in an unknown environment" lies in the lack of knowledge, which includes:
\begin{enumerate}
    \item Lack of \textbf{environmental knowledge}. Although the intelligent agent can observe and interact with the environment, there is no prior knowledge about how the environment will change after interactions. For example, AutoTask can acquire GUI content; however, it lacks knowledge about how the GUI will change after simulating user actions (e.g., clicking a button on the screen).
    \item Lack of \textbf{task knowledge}. The agent does not possess a set of supported tasks or have any predefined understanding related to task semantics or the interpretation of external inputs. For example, we have not provided AutoTask with a predefined set of intents or information about intent parameters. Additionally, we have not provided models or scripts to assist AutoTask in recognizing the intents and parameters within user commands.
    \item Lack of \textbf{execution knowledge}. Given the absence of environmental and task knowledge, the agent lacks the knowledge of how to execute tasks. For AutoTask, this is reflected in the absence of execution scripts (whether provided by developers or end users) for the intent.
\end{enumerate}

To address this challenge, we propose an "explore-learn" strategy that comprises:
\begin{enumerate}
    \item Trial and error: The agent explores the environment and attempts to execute the task. It recovers from errors through backtracking when necessary.
    \item Learn from the environment: The agent accumulates experiences (records of actions and observations) during the exploration of the environment. From these experiences, the agent learns knowledge, enabling it to (a) directly execute intents that have been completed in the past and (b) expedite the exploration when executing unknown intents.
\end{enumerate}

The system design of AutoTask is an application of this strategy to the field of voice assistants, which will be elaborated upon in the next section.
\section{System Design}

\begin{figure}
    \centering
    \includegraphics[width=0.75\linewidth]{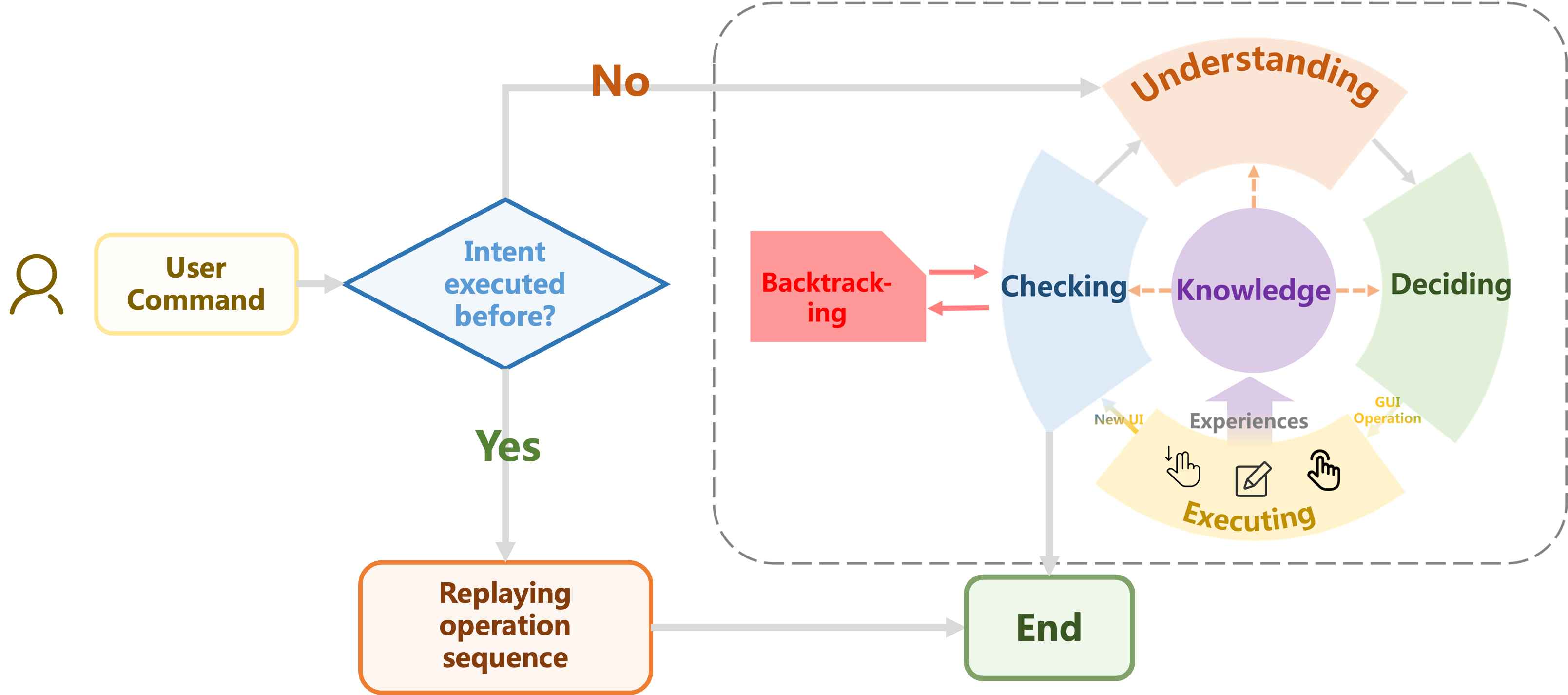}
    \caption{AutoTask's pipeline. It first checks if the intent has been executed. If so, it replays the recorded operation sequence. Otherwise, AutoTask explores the GUI and learns. It understands the GUI and command semantics, deciding the most probable operation. This operation is executed by simulating user actions. AutoTask checks the completed actions with the new GUI content. If an error is detected, it backtracks to undo previous actions. Throughout, AutoTask accumulates experience and summarizes knowledge to enhance its capabilities. The arrows pointing to "Experiences" are omitted in the figure for simplicity.}
    \label{fig: pipeline}
\end{figure}
Figure \ref{fig: pipeline} illustrates the AutoTask pipeline. Upon receiving a user command, AutoTask checks whether the intent expressed in the command has been previously executed. If it has, AutoTask automatically executes the task by replaying the operation sequence, with adjustments based on the current command's parameter values \cite{pan2022automatically}. If the command has not been executed previously or if the replay of the sequence fails, AutoTask enters the "explore-learn" mode, which can be divided into two parts:
\begin{enumerate}
    \item Trail and error, which can be further subdivided into forward exploration and backward backtracking:
    \begin{enumerate}
        \item During the forward exploration, AutoTask selects an optimal operation within the current GUI content (the understanding module and the deciding module), which is then automated by programmatically injecting an event into the GUI (the executing module). After obtaining the resulting GUI content, AutoTask assesses whether the current task is completed and whether the executed operation sequence is correct (the checking module). Based on this assessment, it decides whether to terminate the execution, continue forward exploration, or initiate backward backtracking.
        \item During the backward backtracking, AutoTask undoes the last action (the backtracking module) and evaluates whether the current task is completed and whether the operation sequence is correct (the checking module). AutoTask decides accordingly to terminate the execution, continue backward backtracking, or start forward exploration.
    \end{enumerate}
    \item Learning from the environment, which encompasses: 
    \begin{enumerate}
        \item Accumulating experiences: AutoTask records all its decisions (e.g., outputs of the modules) and GUI observations.
        \item Summarizing knowledge: AutoTask extracts correct knowledge from experiences at appropriate times. The details of AutoTask's knowledge are illustrated in Table \ref{table: knowledge}.
        \item Applying knowledge: AutoTask utilizes its knowledge during task execution to directly execute previously completed intents or expedite the exploration process.
    \end{enumerate}
\end{enumerate}

\begin{table}[ht]
\caption{AutoTask's knowledge}
\label{table: knowledge}
\begin{tabular}{|c|c|l|cc|}
\hline
\multirow{2}{*}{Knowledge} &
  \multirow{2}{*}{Type} &
  \multicolumn{1}{c|}{\multirow{2}{*}{Content}} &
  \multicolumn{2}{c|}{Goal} \\ \cline{4-5} 
 &
   &
  \multicolumn{1}{c|}{} &
  \multicolumn{1}{l|}{\begin{tabular}[c]{@{}l@{}}Directly execute\\ completed intents\end{tabular}} &
  \multicolumn{1}{l|}{\begin{tabular}[c]{@{}l@{}}Expedite Exploration\\ for unknown intents\end{tabular}} \\ \hline
\multirow{2}{*}{\begin{tabular}[c]{@{}c@{}}Environmental\\ Knowledge\end{tabular}} &
  1 &
  \begin{tabular}[c]{@{}l@{}}Contents and transitions \\ within the GUI\\ (used in understanding)\end{tabular} &
  \multicolumn{1}{c|}{-} &
  \checkmark \\ \cline{2-5} 
 &
  2 &
  \begin{tabular}[c]{@{}l@{}}Contents or transitions\\ that do not exist within the GUI\\ (used in deciding and checking)\end{tabular} &
  \multicolumn{1}{c|}{-} &
  \checkmark \\ \hline
\begin{tabular}[c]{@{}c@{}}Task\\ Knowledge\end{tabular} &
  - &
  \begin{tabular}[c]{@{}l@{}}Intents and parameters of commands\\ (used in replaying and understanding)\end{tabular} &
  \multicolumn{1}{c|}{\checkmark} &
  - \\ \hline
\multirow{2}{*}{\begin{tabular}[c]{@{}c@{}}Execution\\ Knowledge\end{tabular}} &
  1 &
  \begin{tabular}[c]{@{}l@{}}Correct operation sequences\\ (used in replaying, deciding, and checking)\end{tabular} &
  \multicolumn{1}{c|}{\checkmark} &
  \checkmark \\ \cline{2-5} 
 &
  2 &
  \begin{tabular}[c]{@{}l@{}}Lessons that prevent execution errors\\ (used in deciding and checking)\end{tabular} &
  \multicolumn{1}{c|}{-} &
  \checkmark \\ \hline
\end{tabular}
\end{table}

\subsection{Learning from the GUI: Summarizing Experiences into Knowledge}
\subsubsection{Experiences of AutoTask}
\begin{figure}
    \centering
    \includegraphics[width=\linewidth]{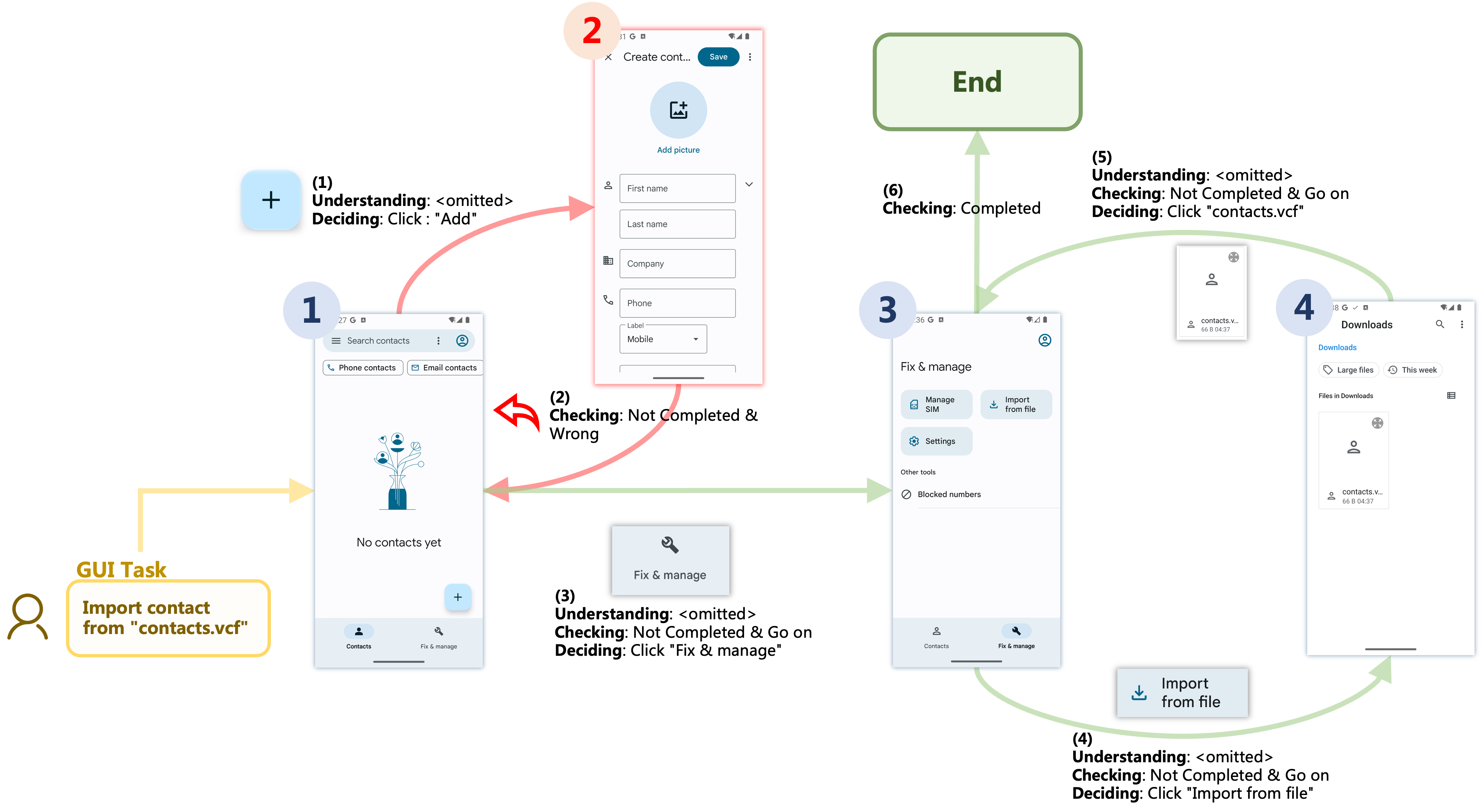}
    \caption{AutoTask's execution experience for the command "Import Contacts from contacts.vcf". (1) AutoTask clicks the "Add" button on Page 1. (2) AutoTask detects an incorrect history operation based on the content of the new page (i.e., Page 2) and backtracks to Page 1, as indicated by the orange arrow. (3) AutoTask determines that the task remains incomplete and selects the "Fix \& Manage" button. (4) AutoTask clicks "Import from file" on the new page. (5) AutoTask selects "contacts.vcf" following the user command. It is a parameter of the intent. (6) AutoTask concludes that the task is completed and finishes the execution.}
    \label{fig: experiences}
\end{figure}

AutoTask automatically records its experiences during runtime, which assists the system in executing the current task and is also summarized into knowledge to support subsequent tasks. As depicted in Figure \ref{fig: experiences}, we represent AutoTask's experiences as a graph, where nodes represent GUI pages, and edges record the results of the modules. Please refer to the corresponding sections for detailed results of each module.

\subsubsection{Environmental knowledge}
AutoTask's environment is the GUI, and its environmental knowledge describes the contents of GUI pages and the transitions between pages. This knowledge can be categorized into the following two types:

\begin{enumerate}
    \item (Type-1) Contents and page transitions of the GUI, which are stored in the form of triplets (S, O, D). S and D represent the GUI pages before and after the operation (represented through HTML, as indicated in Figure \ref{fig: gui-understand}). O = (E, A, P), which denotes elements, actions, and parameters (e.g., the texts for the "text input" actions), respectively; these three components together describe a GUI operation. For example, the Edge (5) (in Figure \ref{fig: experiences}) corresponds to the following triplet: (Page 4, (button labeled "contacts.vcf", click, null), Page 3).
    \item (Type-2) Contents or transitions that do not exist in the GUI. This kind of knowledge is in natural language. For example, one piece of environmental knowledge (Type-2) summarized for Edge (1) (in Figure \ref{fig: experiences}) could be, "By clicking the 'Add' button on the home page, you can only manually adding a single contact. Importing contacts from files is not supported". This knowledge may prevent AutoTask from attempting to click the "Add" button in future commands (e.g., "Import contacts from cloud backup"), thereby expediting the execution of subsequent tasks. 
\end{enumerate}

\subsubsection{Task knowledge}
Task knowledge helps understand the semantics of the tasks. AutoTask's task is to execute natural language commands, the semantics of which are typically described by intents and parameters (also called slots in some works \cite{li2018kite}). A piece of task knowledge comprises (1) an intent name, (2) parameter names, (3) a command corresponding to the intent, and (4) values of the parameters in that command. An intent may appear in multiple pieces of task knowledge, as the commands inside are different. For example, the task knowledge from Figure \ref{fig: experiences} is (1) intent name - "import contacts from file"; (2) parameter name list - "file name"; (3) command - "import contacts from contacts.vcf"; (4) parameter values - "file name = 'contacts.vcf'".

\subsubsection{Execution knowledge}
\begin{figure}
    \centering
    \includegraphics[width=\linewidth]{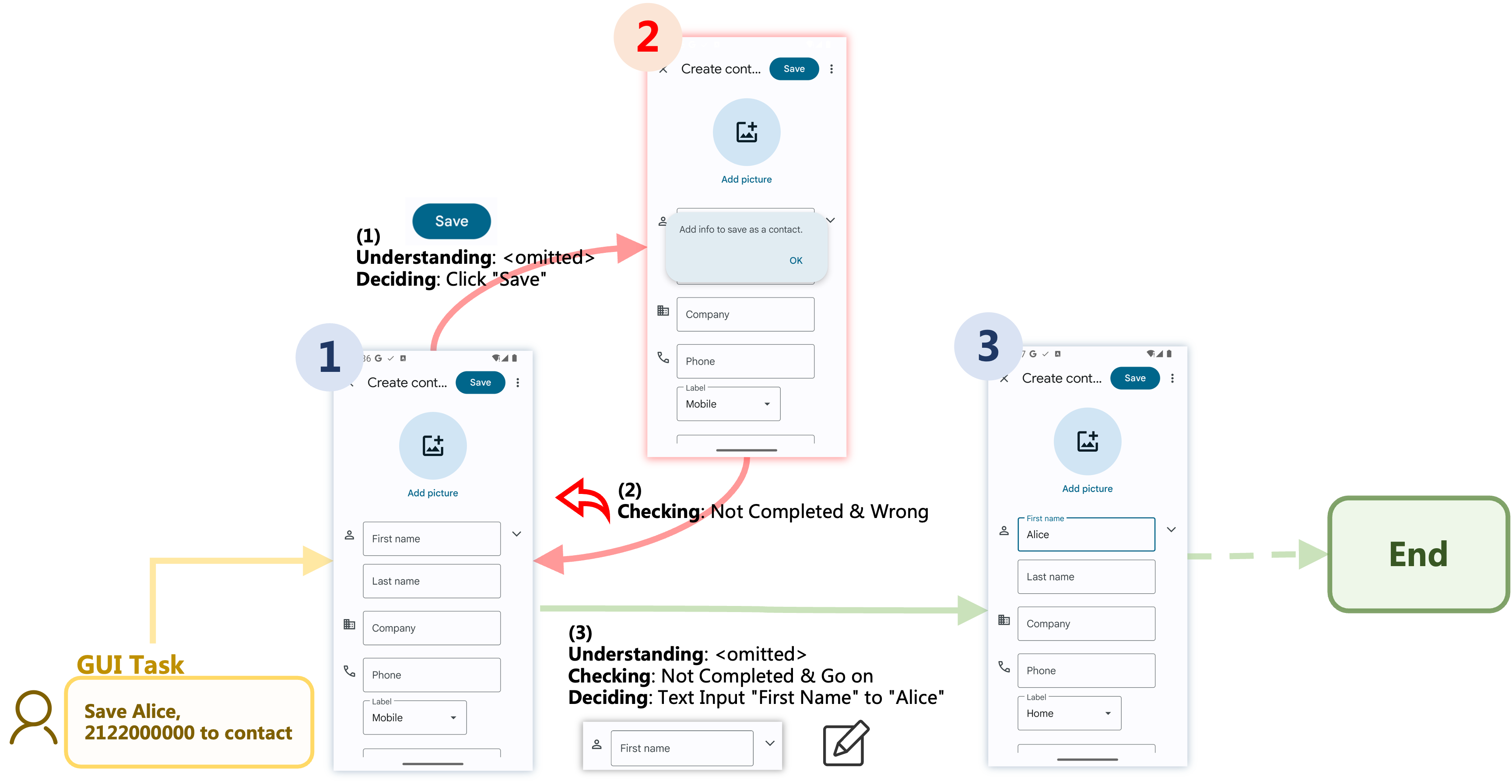}
    \caption{AutoTask's execution experiences for the command "Save Alice, 2122000000 to contact". AutoTask makes a mistake: clicking the Save button (Edge 1) without entering the name and the phone number. This error is corrected through backtracking (Edge 2). AutoTask also summarizes a piece of knowledge to avoid similar mistakes: you should pay attention to the order of steps; actions that may finalize a task (e.g., clicking the Save button) should be performed last.}
    \label{fig: experiences-exec-knowledge}
\end{figure}

Execution knowledge describes how to execute the command in the GUI. It can be categorized into the following two types:
\begin{enumerate}
    \item (Type-1) The correct operation sequence for the command. For example, the first type of execution knowledge for "import contacts from contacts.vcf" is "click 'Fix \& manage', click 'Import from file', click <file name>". The angle brackets (<>) denote parameter values in the command.
    \item (Type-2) Describing how to avoid incorrect operations. This knowledge comprises lessons summarized in natural language, corresponding to errors made by AutoTask during execution. As exemplified in Figure \ref{fig: experiences-exec-knowledge}, AutoTask makes a mistake when executing "save Alice, 2122000000": it clicks "Save" without adding any information about the contact. A possible piece of knowledge summarized from this error could be, "You should pay attention to the order of steps; actions that may finalize a task (e.g., clicking the Save button) should be performed last".
\end{enumerate}

\subsubsection{From experiences to knowledge} \label{sec: from experiences to knowledge}
AutoTask's experiences need to be further summarized into knowledge because some experiences may be redundant or erroneous.

\textbf{Environmental knowledge (Type-1)}: When AutoTask simulates an operation on the GUI and obtains the contents of the resulting page, the experience is immediately transformed into environmental knowledge (Type-1). AutoTask's observations of the environment are always correct; as a result, AutoTask can directly and instantaneously convert the related experiences into knowledge.

\textbf{Execution Knowledge (Type-1)}: AutoTask identifies the shortest path in its experiences that (1) connects the starting point and the endpoint\footnote{The GUI screen where AutoTask thinks the task is completed} and (2) encompasses all parameters. For example, in Figure \ref{fig: experiences}, the path "3-4-5" satisfies the aforementioned criteria. This path is considered the correct operation sequence for the task and is added to the database. Such knowledge is only summarized after the task is completed because AutoTask can only determine the task's endpoint at that time.

\textbf{Task Knowledge}: AutoTask records the command understanding result (to be discussed in \ref{sec: understanding the command}) at the final step of the correct path as a piece of task knowledge. This type of knowledge will be summarized once AutoTask completes the task; otherwise, the command understanding result may be incorrect.

\textbf{Environmental Knowledge (Type-2) \& Execution Knowledge (Type-2)}: The purpose of these two types of knowledge is to prevent errors during task execution. AutoTask compares its experiences with the correct path to identify erroneous steps (e.g., Step 1 in Figure \ref{fig: experiences} \& \ref{fig: experiences-exec-knowledge}). AutoTask categorizes the reasons for errors into two types:

\begin{enumerate}
    \item Lack of environmental knowledge. For example, the error in step 1 of Figure \ref{fig: experiences} occurs because AutoTask does not know whether there will be an "import from file" option after clicking the "Add" button. Since importing from a file is a way to "add" a batch of contacts, AutoTask considers trying the "Add" button worthwhile.
    \item Lack of execution knowledge. For example, the error in step 1 of Figure \ref{fig: experiences-exec-knowledge} happens because AutoTask does not know how to determine the order of operations when multiple GUI actions are related to the command.
\end{enumerate}

AutoTask summarizes a lesson in natural language for each error and utilizes it to avoid future errors. This knowledge can only be summarized after completing the task. Otherwise, AutoTask cannot accurately determine whether a step is correct. We employ LLM to compile this type of knowledge. The prompt will be discussed in section \ref{sec: prompt}. 

\subsection{The Understanding Module}
In the understanding module, AutoTask comprehends the GUI and the user command, the results of which serve as inputs for subsequent modules. This module enables AutoTask to augment information about the environment (i.e., the GUI) and the task (i.e., the user command) with its knowledge.

\subsubsection{Understanding the GUI}\label{sec: understanding the gui}
\begin{figure}
    \centering
    \includegraphics[width=\linewidth]{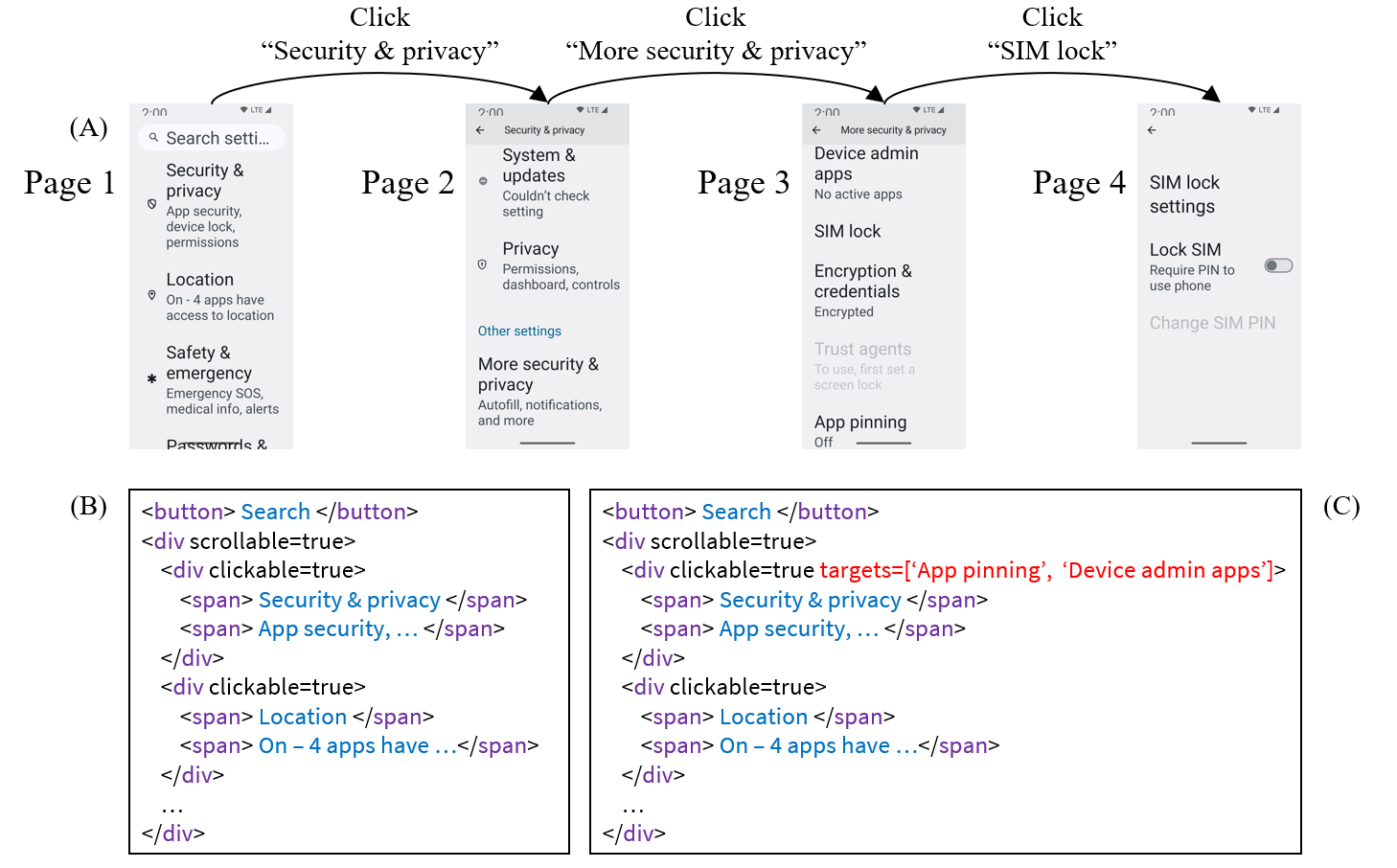}
    \caption{An example result of AutoTask understanding the GUI. (A) The operation sequence for enabling SIM lock. We omit some operations, such as scrolling the screen. AutoTask learns environmental knowledge (Type-1) from the sequence. (B) The HTML representation of Page 1. We omit some GUI elements. (C) The HTML representation after AutoTask understanding the GUI when executing the command "enable app pinning". The understanding result corresponds to the "targets" property of the "Security \& privacy" button.}
    \label{fig: gui-understand}
\end{figure}

The GUI semantics are formed by the contents of GUI pages and the transitions between pages. AutoTask can obtain page data through APIs provided by the operating system (e.g., Android AccessibilityService\footnote{\url{https://developer.android.com/reference/android/accessibilityservice/AccessibilityService}}). However, the incompleteness of GUI semantics arises because the contents after GUI operations cannot be foreseen. AutoTask addresses this issue by querying environmental knowledge to infer the elements "hidden" behind the current GUI elements. This process can be divided into two steps:
\begin{enumerate}
    \item AutoTask retrieves elements from the environmental knowledge (Type-1) that can be reached through one or multiple operations starting from the current GUI elements.
    \item To filter out irrelevant elements, these elements are transformed into vectors, and their similarities to the user command are computed (details are discussed in section \ref{sec: similarity}). The semantic understanding result includes only elements with similarities exceeding a threshold. As exemplified in Figure \ref{fig: gui-understand} (C), the button "App pinning" is very related to the user command "enable app pinning" and is reachable by operating an element on the current GUI (clicking "Security \& privacy" and then clicking "More security \& privacy"). As a result, it is added to the "target" property of the button "Security \& privacy".
\end{enumerate}

The GUI semantics effectively guide AutoTask in selecting the correct operations. As illustrated in Figure \ref{fig: gui-understand}, AutoTask successfully executes the command "enable SIM lock". This is not challenging since SIM lock and security are highly semantically related. AutoTask also learns GUI-related knowledge during execution. Next, AutoTask executes the command "enable App pinning". Without relevant GUI knowledge, executing this command is challenging: AutoTask may blindly attempt to click on different buttons on the Android Settings homepage, such as "Application", "Display", and "Safety". However, GUI knowledge can assist AutoTask in directly choosing "Security \& privacy" without additional explorations. During GUI understanding, AutoTask discovers a high semantic similarity between the "App pinning" button and the user command, and there exists an operation sequence from "Security \& privacy" to "App pinning".

\subsubsection{Understanding the command} \label{sec: understanding the command}
During the process of understanding command semantics, AutoTask generates a natural language phrase that describes the intent conveyed in the user command. It also detects the parameters and their values from the command. Command understanding needs to be performed in each iteration of AutoTask because the understanding results may be updated as experiences accumulate \cite{pan2022automatically}.

The command semantics can guide AutoTask in taking correct operations in the GUI. For example, in the command "Save Alice, 2122000000 to contact" (Figure \ref{fig: experiences-exec-knowledge}), if AutoTask realizes that "Alice" and "2122000000" are parameters for "create a new contact", it will use the two parameters during task execution, that is, entering them into text boxes\footnote{Another way to use a parameter is to select an item with corresponding text in a list \cite{pan2022automatically}.}. The results of command understanding may be stored as task knowledge (as already discussed in \ref{sec: from experiences to knowledge}), which can assist AutoTask in comprehending subsequent commands.

AutoTask employs task knowledge and an LLM to understand the semantics of commands. We calculate the semantic similarities between the historical commands in the task knowledge and the current user command (details will be discussed in \ref{sec: similarity}). Historical commands with similarities greater than a threshold will be selected as examples. The LLM will utilize these examples, the executed operation sequence, and current GUI contents to calculate the command understanding results. The prompts will be discussed in \ref{sec: prompt}.

\subsection{The Deciding Module}\label{sec: decision module}
\begin{table}[ht]
\caption{Types of operations supported by AutoTask. Click, text input, and scroll forward are used in the deciding module, while the other three are used in the backtracking module.}
\label{table: operations}
\begin{tabular}{|c|c|c|c|}
\hline
Action          & GUI Elements         & Parameters   & Usage                                                                                                   \\ \hline
Click           & Clickable elements  & N/A          & \multirow{3}{*}{The Deciding Module}                                                                    \\ \cline{1-3}
Text Input      & Editable elements   & Text content &                                                                                                         \\ \cline{1-3}
Scroll Forward  & Scrollable elements & N/A          &                                                                                                         \\ \hline
Navigate Up     & N/A                 & N/A          & \begin{tabular}[c]{@{}c@{}}The Backtracking Module\\ (to undo a click operation)\end{tabular}           \\ \hline
Clear Text      & Editable elements   & N/A          & \begin{tabular}[c]{@{}c@{}}The Backtracking Module\\ (to undo a text input operation)\end{tabular}      \\ \hline
Scroll Backward & Scrollable elements & N/A          & \begin{tabular}[c]{@{}c@{}}The Backtracking Module\\ (to undo a scroll forward operation)\end{tabular} \\ \hline
\end{tabular}
\end{table}

In the deciding module, AutoTask calculates the most possible operation in the current GUI to complete the user command. AutoTask identifies all operations\footnote{The parameter for text input will be determined later} available in the GUI. Table \ref{table: operations} provides an overview of the types of supported operations. Subsequently, AutoTask assigns scores to these operations, with higher scores indicating a greater possibility of being the next operation. Each operation's score is calculated based on a basic score and a penalty factor: $score = basic\_score / (1 + penalty)$:
\begin{enumerate}
    \item Basic Score (1.0 - 8.0), which is calculated by adding the following two components together:
    \begin{enumerate}
        \item Likert scale (1.0 - 7.0). We employ an LLM to assess the relevance of the operations to the user's command. We use a 7-point Likert scale, where 1 indicates extremely low relevance, and 7 denotes very high relevance. To assist the LLM in calculating the scores, we retrieve relevant environmental knowledge (Type-2) and execution knowledge (Type-1 \& 2) from the knowledge base, as discussed in \ref{sec: similarity}. Further details about the prompt will be provided in \ref{sec: prompt}.
        \item Tie-breaking score (0.0 - 1.0). We calculate the semantic similarities between operations and tasks (please refer to section \ref{sec: similarity} for more details) as the tie-breaking scores. The tie-breaking scores increase the differentiation between different operations. While Likert scales are typically effective at identifying the most relevant operation, they may lack granularity when scoring less relevant options. The tie-breaking scores prevent operations from receiving identical scores, thereby avoiding AutoTask being reduced to brute-force searching.
    \end{enumerate}
    \item Penalty factor (0.0 - positive infinity). It is important to note that penalizing an operation does not necessarily mean that the final score of the element will not be the highest. For example, when an operation is penalized by the checking module, AutoTask may attempt other operations and find that these operations are even less relevant to the user command. In this case, AutoTask may retry the penalized operation. The penalty factor consists of two components:
    \begin{enumerate}
        \item Repetition penalty, used to penalize operations that have already appeared in the executed operation sequence\footnote{Note that a combination of an action, a GUI element, and a parameter describes an operation. A repetitive operation implies that AutoTask has arrived at the current GUI screen.}. The repetition penalty is fixed at 10.
        \item Backtracking penalty, used to penalize operations considered incorrect by the checking module. The backtracking penalty is initialized at 0 and can be updated by the checking module (see \ref{sec: checking module} for details).
    \end{enumerate}
\end{enumerate}

AutoTask selects the operation with the highest score as the next to be executed. If the current operation is text input, AutoTask utilizes an LLM to calculate the text content. Please refer to \ref{sec: prompt} for more details about the prompt.

\subsection{The Executing Module \& the Backtracking Module}
In the executing module, we utilize the accessibility API to inject operations into the GUI based on the results from the deciding module. Conversely, the backtracking module injects interaction events (e.g., scrolling backward) to undo previous operations (e.g., scrolling forward, as indicated in Table \ref{table: operations}). Both modules retrieve the GUI hierarchy after injecting events, which will be converted into HTML format (as shown in Figure \ref{fig: gui-understand}(B)) and used by other modules. Operations may only cause some minor localized changes in the GUI, so we compare the GUI pages before and after the operations to identify newly appeared elements, which are then marked with a boolean property named "new" (as shown in Figure \ref{fig: appendix-context} in the Appendix).

We remove elements that meet both of the following two criteria from the GUI hierarchy:
\begin{enumerate}
    \item Low interaction importance. This criterion applies when the element itself and its descendant elements (if any) cannot be interacted with and do not contain text or descriptions.
    \item Low layout importance. This criterion applies when the element has no sibling elements or all sibling elements are considered to have "low interaction importance".
\end{enumerate}

\subsection{The Checking Module}\label{sec: checking module}
After AutoTask performs GUI operations (both in the executing module and the backtracking module), the checking module conducts two checks on the completed operation sequence: completeness and correctness.

\subsubsection{completeness check}
The checking module employs an LLM to determine whether the current task is completed. This check is also performed during the backtracking process to address "overshoot" issues, where unnecessary operations are executed after task completion. For more prompt details, please refer to \ref{sec: prompt}. The checking module also considers the task completed when the number of executed steps exceeds a threshold (set to 20 in our implementation).

When AutoTask considers the task as completed, we present the user with a list describing the shortest execution path (as discussed in \ref{sec: from experiences to knowledge}). Each item in the list includes (1) a screenshot, (2) a rectangular bounding box used to highlight the operated element in the screenshot, and (3) a text description of the operation, e.g., "Text input: Alice". The user can choose one of the following options:
\begin{enumerate}
    \item Confirming the correctness of the execution process. The system summarizes the knowledge and terminates.
    \item Confirming that the task is not yet completed. The system continues running and starts the correctness check. The current page will not be considered as the endpoint for the current command.
    \item Forcing termination. The system stops running directly without knowledge summarization. Note that the first type of environmental knowledge is still summarized and accumulated.
    \item Ignoring (default\footnote{In the evaluation study, we assumed that users would select this option.}). If the step threshold is exceeded, AutoTask stops running without knowledge summarization. Otherwise, the system summarizes knowledge and terminates.
\end{enumerate}

\subsubsection{correctness check}
If a task is not completed, AutoTask checks whether the last step\footnote{Backtracking steps or the steps being undone will not be considered as last steps. For example, if "A-B-C" forms an operation sequence and AutoTask uses operation D to undo operation C, then, even though the sequence is "A-B-C-D", we regard B as the last step.} currently being executed is correct, i.e., whether AutoTask can continue to fulfill the user's instruction. The essence of a correctness check is checking the correctness of the deciding module with more experiences and knowledge accumulated from the GUI. For example, in Figure \ref{fig: experiences}, AutoTask clicks the "Add" button to import contacts from the file. However, after simulating user interaction and obtaining new GUI contents, the checking module can discover that the result of the deciding module is erroneous.

AutoTask applies an LLM to conduct the correctness check. Please refer to \ref{sec: prompt} for the detailed prompt of LLM. It is worth noting that AutoTask takes into account the backtracking penalty of the last step. If an operation has a high backtracking penalty but is still executed, the possibility of other operations may be lower. In such a case, the checking module should be more tolerant and consider it correct, providing an opportunity for further exploration.

If the last operation is considered to be incorrect, the checking module will calculate a penalty (0-9) to describe the severity of the error: 0 indicates that the error in the last operation is due to preceding steps already being incorrect; 9 indicates a very serious error in the last operation itself. This penalty will be accumulated into the current backtracking penalty of the last operation. Consequently, the backtracking penalty of an operation may be greater than 9, which indicates that it has been rejected by the checking module several times. We use LLM to calculate the penalty, with details of its prompt discussed in \ref{sec: prompt}.
\section{Implementation}
In this section, we describe how AutoTask utilizes LLMs for computational purposes. Please refer to the Appendix for more detailed examples.

\subsection{AutoTask's context}
AutoTask's context describes its state and is widely used throughout the computational process, as shown in Figure \ref{fig: appendix-context} in the Appendix. It encompasses (1) the user command, (2) the executed operation sequence, and (3) the current GUI contents represented in HTML. The GUI content is augmented with environmental knowledge (Type-1), as discussed in \ref{sec: understanding the gui}. (4) the latest semantic comprehension result of the instruction (if any).

\subsection{Embedding \& Similarity: Choosing One or More Answers from Several Candidates}\label{sec: similarity}
AutoTask utilizes the Embedding \& Similarity approach to select one or multiple answers from several candidates for a given question. Both the question and the candidates are transformed into vectors using the embedding API provided by OpenAI. We regard each candidate's cosine similarity with the question as its score. A higher score indicates that the candidate is more likely to be the correct answer. Table \ref{table: es usage} summarizes the usage of this method. It is worth noting that the description of elements includes their surrounding elements, as they may exhibit strong semantic relevance.

\begin{table}[ht]
\caption{Usage of the Embedding \& Similarity approach. To prevent the table from becoming too wide, we use "env." as an abbreviation for "environmental" and "exe." as an abbreviation for "execution".}
\label{table: es usage}
\begin{tabular}{|l|l|l|}
\hline
Purpose &
  Question &
  Candidate Answers \\ \hline
\begin{tabular}[c]{@{}l@{}}Retrieve elements from\\ env. knowledge (Type-1)\\ (\ref{sec: understanding the gui})\end{tabular} &
  \begin{tabular}[c]{@{}l@{}}(1) context\\ (2) "what element is related"\end{tabular} &
  \begin{tabular}[c]{@{}l@{}}Elements\\ in env. knowledge (Type-1)\end{tabular} \\ \hline
\begin{tabular}[c]{@{}l@{}}Filter relevant pieces of\\ task knowledge\\ (\ref{sec: understanding the gui})\end{tabular} &
  \begin{tabular}[c]{@{}l@{}}(1) context\\ (2) "what task is similar to the command"\end{tabular} &
  \begin{tabular}[c]{@{}l@{}}Tasks and its semantics\\ from task knowledge\end{tabular} \\ \hline
\begin{tabular}[c]{@{}l@{}}Filter relevant pieces of\\ env. knowledge and\\ exe. knowledge\\ (\ref{sec: decision module} \& \ref{sec: checking module})\end{tabular} &
  \begin{tabular}[c]{@{}l@{}}(1) context\\ (2) "what knowledge is related"\end{tabular} &
  \begin{tabular}[c]{@{}l@{}}Env. knowledge (Type-2)\\ Exe. knowledge (Type-1 \& 2)\end{tabular} \\ \hline
\begin{tabular}[c]{@{}l@{}}Compute tie-breaking score\\ (\ref{sec: decision module})\end{tabular} &
  \begin{tabular}[c]{@{}l@{}}(1) context\\ (2) "what element should be operated"\end{tabular} &
  Current GUI elements \\ \hline
\end{tabular}
\end{table}

\subsection{Text Completion: Answering Questions}\label{sec: prompt}
AutoTask utilizes the text completion approach to generate an answer for a given question. The question is passed as part of the prompt to the LLM (gpt-4), and the response generated by the LLM serves as the answer. Table \ref{table: tc usage} summarizes the usage of this method. The output template specifies the JSON format the LLM response should adhere to, which AutoTask can parse easily.

\begin{table}[ht]
\caption{Usage of the Text Completion method}
\label{table: tc usage}
\begin{tabular}{|c|c|l|}
\hline
Module &
  Purpose (question) &
  \multicolumn{1}{c|}{Prompt Composition} \\ \hline
Understanding &
  Understand command semantics &
  \begin{tabular}[c]{@{}l@{}}(1) Purpose\\ (2) Task knowledge\\ (3) Context\\ (4) Output template\end{tabular} \\ \hline
\multirow{2}{*}{Deciding} &
  Calculate the Likert scale &
  \begin{tabular}[c]{@{}l@{}}(1) Purpose\\ (2) Execution knowledge\\ (3) Context\\ (4) Output template\end{tabular} \\ \cline{2-3} 
 &
  \begin{tabular}[c]{@{}c@{}}Calculate parameter for\\ text input\end{tabular} &
  \begin{tabular}[c]{@{}l@{}}(1) Purpose\\ (2) Context\\ (3) Textbox to be edited\\ (4) Output template\end{tabular} \\ \hline
Check &
  \begin{tabular}[c]{@{}c@{}}Checking completeness\\ Check correctness\\ Calculate penalty\end{tabular} &
  \begin{tabular}[c]{@{}l@{}}(1) Purpose\\ (2) Execution knowledge\\ (3) GUI before last operation\\ (4) Context\\ (5) Output template\end{tabular} \\ \hline
N/A &
  Summary Knowledge &
  \begin{tabular}[c]{@{}l@{}}(1) Purpose\\ (2) AutoTask experiences\\ (3) The erroneous step \\ (4) The ground truth\\ (5) Output template\end{tabular} \\ \hline
\end{tabular}
\end{table}

\section{Evaluation Study}
The evaluation study has two goals: (1) to validate that AutoTask can correctly execute user instructions without predefined knowledge, and (2) to validate that knowledge accumulation can effectively accelerate AutoTask's execution of user commands. We did not evaluate AutoTask's feasibility regarding executing intents that have been carried out before. In such cases, AutoTask only replays the recorded operation sequences (with parameter adjustments), and the feasibility has already been validated in previous works \cite{pan2022automatically}.

\subsection{Apparatus}
We conducted the evaluation on an Android virtual machine (Pixel\_XL running Android 11). No modifications were made to the virtual machine or system, and AutoTask can run on commercial physical devices.

\subsection{Tasks}
We validated the capabilities of AutoTask on the following datasets:
\begin{enumerate}
    \item PixelHelp \cite{li2020mapping}. Consisting of natural language commands and their corresponding operation sequences, PixelHelp was revised for this study. Due to system and application upgrades, some outdated instructions were manually removed, leaving a total of 67 instructions.
    \item UGIF \cite{venkatesh2022ugif}. This dataset also comprises natural language commands and their corresponding operation sequences. We concentrated on instructions related to Android Settings, as knowledge accumulation is more pronounced within the same application. We used these tasks to assess the impact of knowledge accumulation on AutoTask's performance. Similar to PixelHelp, outdated instructions were removed, resulting in 100 remaining instructions.
\end{enumerate}

We made adjustments to the commands in the datasets, including:
\begin{enumerate}
    \item Changing the tone from inquiry to command. Both datasets were compiled from tutorials on the internet, where instructions were mostly presented in the form of inquiries. We modified the instructions to align them with how users typically interact with voice assistants. For example, we changed "how to turn off WiFi" to "turn off WiFi".
    \item Supplementing missing parameters. Some commands lacked parameters and were not executable. We added parameters to these instructions. For example, we changed "how to delete a Google account" to "delete the Google account named Alice".
\end{enumerate}

\subsection{Metrics}\ 

\textbf{Success rate}, i.e., the ratio of the number of successfully completed tasks to the total number of tasks.

% \textbf{Completion determination precision}, i.e., the ratio of the number of successfully completed tasks to the number of tasks identified as completed by the system. Some tasks may not have been completed in reality, but the system may perceive them as completed. This metric measures the accuracy of the system's determination of task completion.

The \textbf{step accuracy} of a task, i.e., the ratio of the number of correct steps to the minimum number of steps required to complete the task. The correct steps are defined as the longest subsequence (instead of substring) of the actually executed steps that satisfy the following criterion: the minimum steps needed to complete the task start with the subsequence. For completed tasks, this metric is always 1. For tasks that were not successfully completed, this metric indicates the proximity of AutoTask to successful completion.

The \textbf{step redundancy rate} of a task, i.e., the ratio of the difference between the number of executed steps and the number of correct steps to the number of executed steps. This metric evaluates the system's efficiency. Tasks that succeeded with a step redundancy rate of 0 are referred to as "tasks completed without redundancy".

\textbf{Non-redundant completion rate}, i.e., the ratio of the number of tasks completed without redundancy to the total number of tasks.

\subsection{Baseline}
We utilized the LLM approach proposed by Android in the Wild (AITW) \cite{rawles2023android} as the baseline. Similar to AutoTask, this approach employs an LLM to automate user instructions without requiring any configuration or modifications, making it applicable to arbitrary GUI intents. However, the baseline solution lacks an explicit self-checking and backtracking mechanism, although it can undo previous actions by performing certain actions (e.g., clicking the back button). Furthermore, it does not summarize and accumulate knowledge from the execution process. Note that in the original baseline solution, the prompt used only included information about the most recent five operations. We adjusted it to include the complete operation sequence to ensure consistency with AutoTask. When the number of execution steps in the baseline approach exceeds 20, we also forcibly terminate it.

\subsection{Procedure}
AutoTask and the baseline are executed in the same order for the commands from the two datasets (shuffled beforehand). After completing each task (normal completion or exceeding the maximum number of steps), the next instruction is automatically executed. Throughout this process, AutoTask accumulates knowledge when a task ends successfully but does not utilize knowledge derived from other tasks\footnote{AutoTask still utilizes the first type of environment knowledge accumulated during the execution of the current task.}. After completing all tasks, experimenters manually check whether each task has been executed correctly.

We categorized the tasks in UGIF into three types based on AutoTask's execution results: (Type-1) tasks that AutoTask can complete without redundancy; (Type-2) tasks that AutoTask can complete with redundancy; (Type-3) tasks that AutoTask is unable to complete. Tasks of Type-1 and 2 are collectively referred to as Type-A tasks; AutoTask has already summarized knowledge for Type-A tasks in Phase 1. Tasks of Type-2 and 3 are collectively referred to as Type-B tasks; there is room for improvement in AutoTask's performance on Type-B tasks. 

We then evaluate the improvement of AutoTask's performance with the accumulated knowledge. For each Type-B task, we randomly select a certain number of tasks from Type-A tasks\footnote{We guarantee that the selected Type-A tasks do not include the Type-B task to be tested.}. We evaluate the performance of AutoTask after accumulating the knowledge from these tasks. We repeat the aforementioned random selection process ten times and compute the average results to mitigate the influence of random noise. To explore the impact of the amount of knowledge on AutoTask's performance, we repeated the process several times with different percentages of the selected Type-A tasks: 20\%, 40\%, 60\%, 80\%, and 100\%.

\subsection{Results}

\subsubsection{Success rate}
The accuracy of AutoTask in PixelHelp is 91.2\% (6 errors in 67 commands) and that in UGIF is 93.0\% (7 errors in 100 commands). The two metrics for the baseline are 52.2\% (32 errors in 67 commands) and 67.0\% (33 errors in 100 commands), respectively. The chi-square test (p < 0.001) proved that AutoTask significantly outperformed the baseline in both datasets. We identified the following two reasons:
\begin{enumerate}
    \item AutoTask demonstrates excellent capability in detecting task completion, with precision at 99.4\% and recall at 100\%. Although the baseline achieves 100\% accuracy in verifying task completion, its recall rate is relatively low: 36.1\% (15 tasks from PixelHelp and 11 tasks from UGIF) of the errors are due to "overshoot", that is, the baseline executes unnecessary steps after the tasks had already completed.
    \item AutoTask exhibits higher accuracy in its behavior on the GUI (results from the deciding and checking modules). The step accuracy of AutoTask is 93.5\% (PixelHelp: 93.6\%, UGIF: 93.4\%), whereas the baseline stands at 82.9\% (PixelHelp: 77.4\%, UGIF: 87.2\% ). %When considering only failed tasks, these metrics are 16.7\% (PixelHelp: 27.0\%, UGIF: 6.12\%) and XX\% (PixelHelp: 76.5\%, UGIF: 61.2\%), respectively. It's important to note that the step accuracy of the overshoot tasks is always 1. Excluding these tasks further reduces the baseline's step accuracy within erroneous tasks to xx\%.
\end{enumerate}

We analyzed tasks that AutoTask did not complete correctly and categorized the reasons for errors into three main types:
\begin{enumerate}
    \item AutoTask failed to properly ground instructions to the GUI, leading to blind attempts on the interface. A typical example is the command "check my chromebook if any", where AutoTask did not recognize "chromebook" as a connected device and instead kept clicking irrelevant buttons such as "Display" and "System". Three tasks failed due to this reason.
    \item AutoTask was misled by information in the command, continuously trying to accomplish irrelevant tasks. For instance, with the command 'lock screen when app unpinning', AutoTask focused on what it perceived as the keyword "lock" and repeatedly tried to add a personal identification number (PIN) to the phone. Nine tasks failed due to this misunderstanding.
    \item AutoTask mistakenly believed the task was completed. This occurred once in our experiments with the command "show system applications". AutoTask successfully navigated to the "Installed Applications" page and thought the task was finished. However, it was expected to click the "More Options" button and then select the "Show System" option.
\end{enumerate}

\subsubsection{Redundancy}
The step redundancy rates of AutoTask across both datasets are 8.54\% and 8.96\%, significantly lower than the corresponding baseline results (46.0\%, p<0.001; 32.0\%, p<0.001). AutoTask necessitates backtracking in only 12 (PixelHelp: 7, UGIF: 5) (7.14\%) tasks, with an average backtrack count of 2.83 (min=1, max=8, sd = 5.18) within these tasks. This indicates that AutoTask requires minimal backtracking to accomplish tasks.

\begin{table}[ht]

\label{table: res}
\begin{tabular}{cccccc}
\toprule
Model      & Dataset &Success Rate        & Step Accuracy   & Step Redundancy Rate & Non-redundant Completion Rate  \\
\midrule
\multirow{2}{*}{Baseline}     & PixelHelp & 52.2\% & 77.4\% & 46.0\% &  65.3\%\\\cline{2-6} 
 & UGIF & 67.0\% & 87.2\% & 32.0\% & 78.4\%\\\cline{1-6}
\multirow{2}{*}{AutoTask} & PixelHelp & 91.2\% & 93.6\% & 8.54\% & 89.7\%\\\cline{2-6}
 & UGIF & \textbf{93.0\%} & 93.4\% & 8.96\% & 95.0\%\\
\bottomrule
\end{tabular}
\caption{Evaluation Results of Baseline LLM (AITW) and AutoTask on PixelHelp dataset and UGIF dataset.}
\end{table}

\subsubsection{Performance improvement through knowledge accumulation}
\begin{figure}
    \centering
    \includegraphics[width=\linewidth]{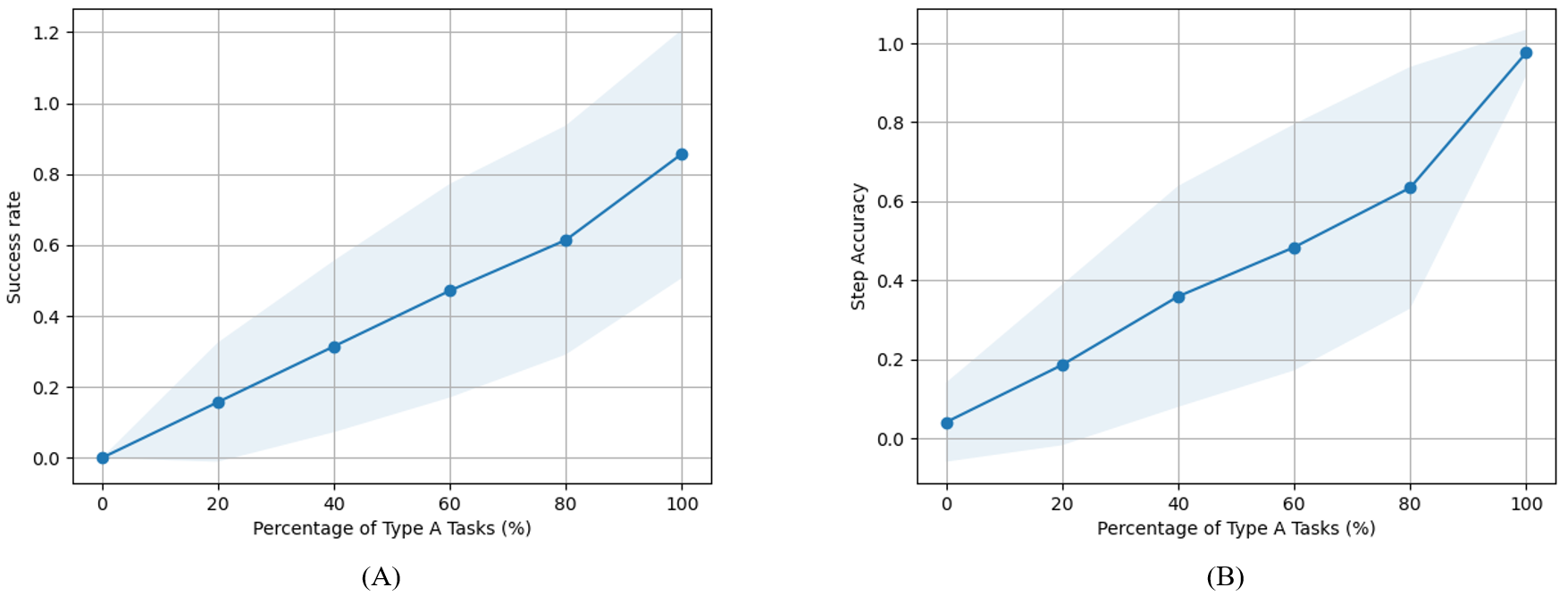}
    \caption{The success rate (A) and step accuracy (B) of Type-3 tasks increase as the percentage of Type-A tasks from which AutoTask accumulates knowledge increases. The shadow represents the standard deviation.}
    \label{fig: T3-success}
\end{figure}
The quantities of the three task types in UGIF are 88, 5, and 7, respectively. Figure x illustrates how the success rate and the step accuracy increase with the accumulation of knowledge. When AutoTask learns all the knowledge, these metrics can reach 85.7\% and 97.6\%, respectively. Only one Type-3 task cannot be completed even after accumulating all the knowledge.

\begin{figure}
    \centering
    \includegraphics[width=\linewidth]{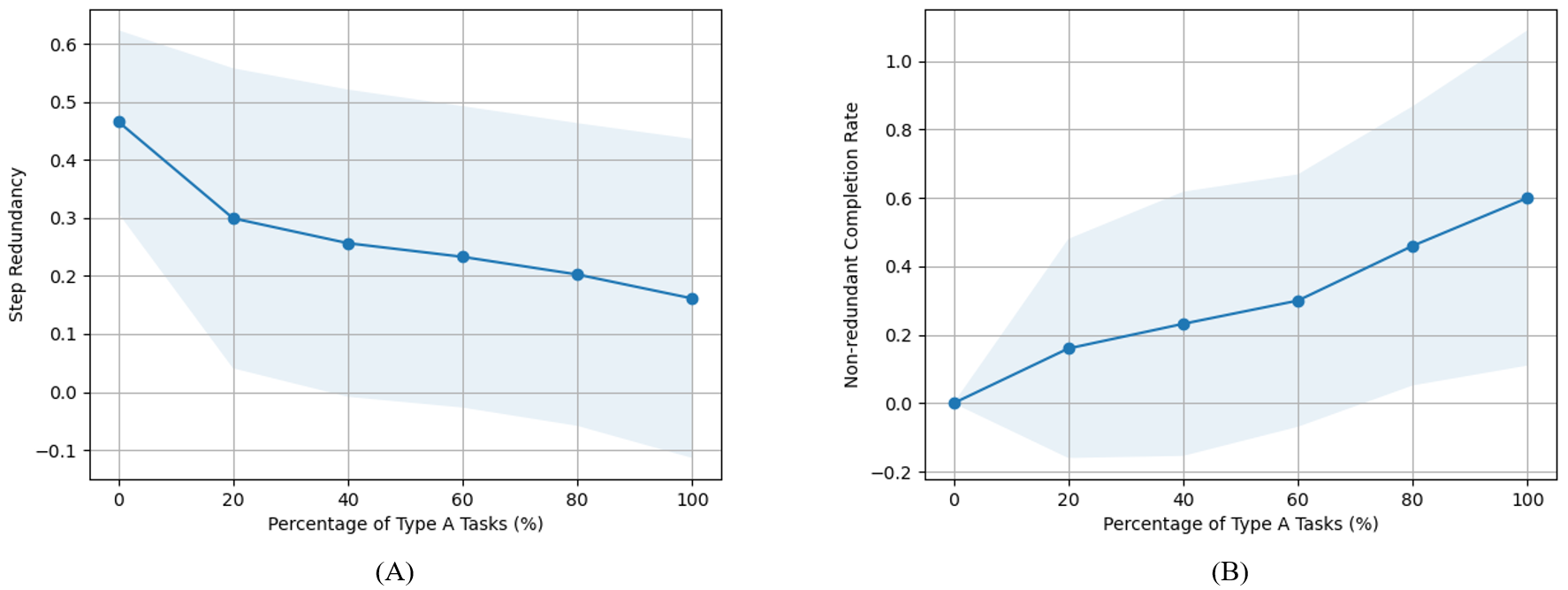}
    \caption{The step redundancy rate (A) and non-redundant completion rate (B) of Type-2 tasks change as the percentage of Type-A tasks from which AutoTask accumulates knowledge increases. The shadow represents the standard deviation.}
    \label{fig: T2-redundancy}
\end{figure}

\begin{figure}
    \centering
    \includegraphics[width=\linewidth]{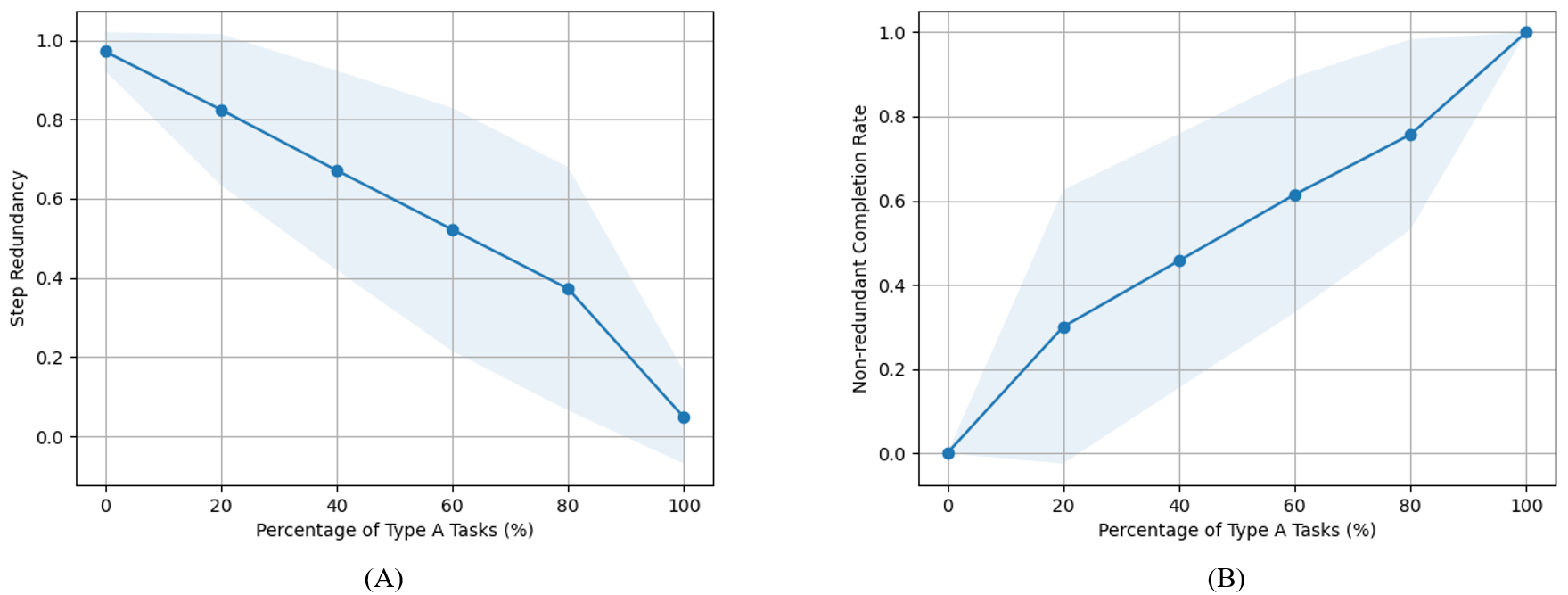}
    \caption{The step redundancy rate (A) and non-redundant completion rate (B) of Type-3 tasks change as the percentage of Type-A tasks from which AutoTask accumulates knowledge increases. The shadow represents the standard deviation.}
    \label{fig: T3-redundancy}
\end{figure}

Furthermore, knowledge accumulation effectively improves the efficiency of task completion. Figure \ref{fig: T2-redundancy} \& \ref{fig: T3-redundancy}demonstrates how the average step redundancy rate and non-redundant completion rate for Type-2 and Type-3 tasks change as the accumulated knowledge increases. Upon acquiring all available knowledge, the step redundancy rates for these two task types decrease from 46.7\% to 16.1\% and from 97.1\% to 4.76\%, respectively, with only 3 tasks (2 Type-2, 1 Type-3) still requiring backtracking.

\section{Discussion}
\subsection{Generalization of Knowledge: Across Versions and Applications}

\begin{figure}
    \centering
    \includegraphics[width=1\linewidth]{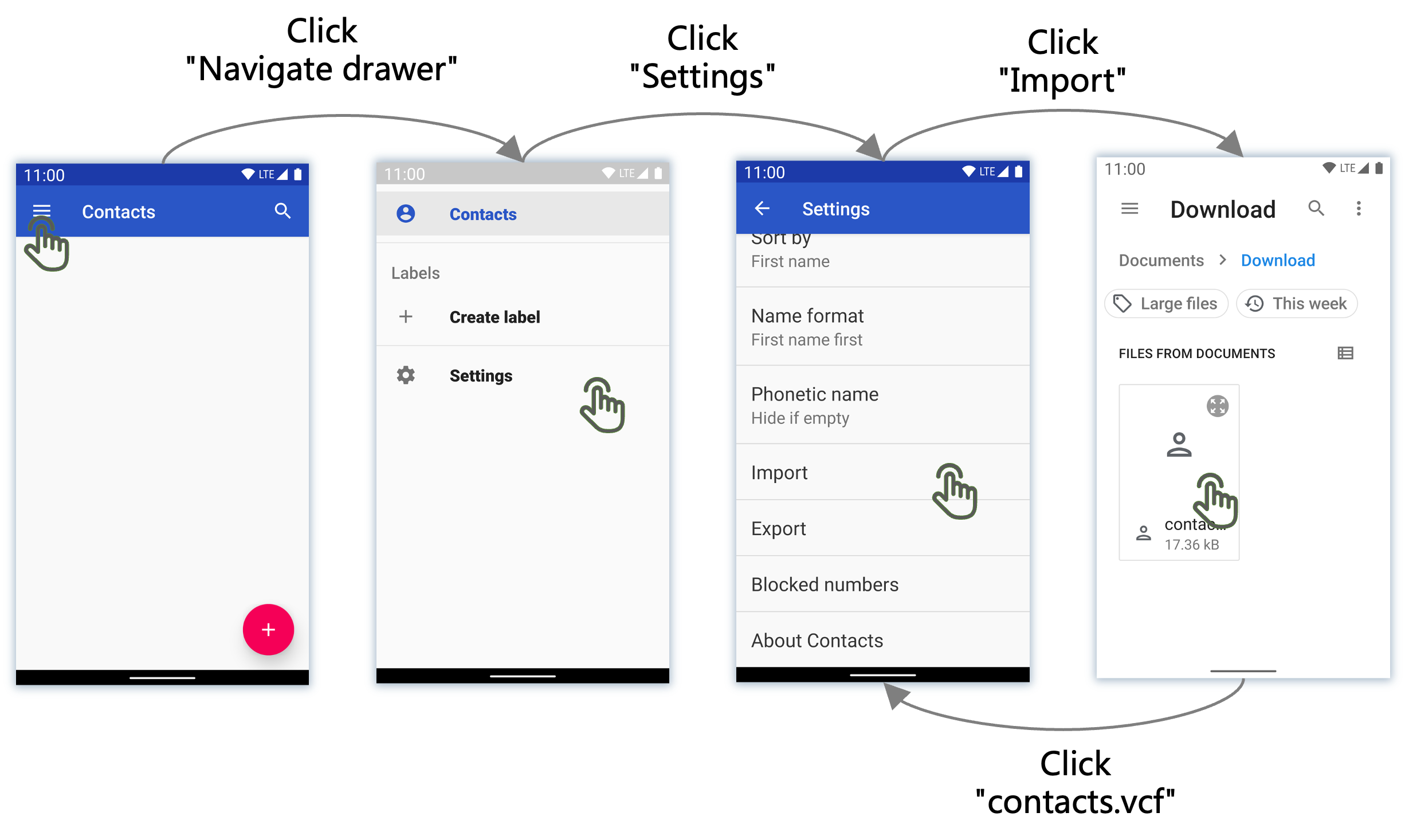}
    \caption{How AutoTask executes "Import contacts from contacts.vcf" in version 1.7.31 of the Contacts application. Note that the location of this functionality has changed compared to Figure \ref{fig: experiences}.}
    \label{fig: old contacts}
\end{figure}

While the study evaluated the performance improvement brought about by knowledge accumulation within the same application, this knowledge can be applied across different versions of the same application or even across applications with similar functionalities (e.g., iMessage vs. WhatsApp). These applications follow similar design principles and semantics \cite{10.1145/3242587.3242650}. For instance, Figure \ref{fig: old contacts} depicts an earlier version of a contact application (version 1.7.31). In contrast to Figure \ref{fig: experiences}, its home page does not contain "Fix \& manage", and "Import from file" is hidden under "Settings". However, AutoTask can still benefit from the knowledge it has summarized. For example, it will refrain from attempting to click the "Add" button, thereby expediting the execution of the command.

The generalization of knowledge across versions and different applications also carries some risks, as the accuracy of such knowledge cannot be guaranteed. For example, if AutoTask first imports contacts from a file in version 1.7.31 of the Contacts application, it may summarize the knowledge that this functionality can be found by clicking the "Settings" button. However, when it attempts the same task in version 4.8.17 of the Contacts application based on this knowledge, it might experience decreased efficiency, even though it can still complete the task correctly through backtracking. This is because the functionality of importing contacts from a file has been relocated in the new version. One solution is to estimate the confidence level of its knowledge. While this goes beyond the scope of this paper, it is a promising direction for future research.

\subsection{The Generalization of AutoTask: to Other Devices and Tasks}

While we implemented AutoTask on Android smartphones and conducted evaluation experiments, AutoTask can generalize to other devices and platforms (e.g., web browsers \cite{leshed2008coscripter, little2007koala}) as long as they provide APIs, with which AutoTask can (1) retrieve the current contents on the GUI and (2) simulate user interactions in the GUI. AutoTask may also generalize to non-graphical interfaces, such as command line interfaces \cite{liu2023agentbench}, to reduce the learning curve and interaction burdens. Besides, AutoTask can be viewed as a proxy for existing GUIs \cite{10.1145/3610929, zhang2017interaction}, and its interaction modality is not limited to voice interaction. GUI mapping \cite{10.1145/3447993.3483245} (e.g., mapping a smartphone GUI to a smartwatch GUI \cite{10.1145/3379503.3403564}) is a typical example. Developers only need to be concerned with the visual mapping rules \cite{chen2018ui} instead of the execution logic of the applications. AutoTask can automatically operate the original GUI (e.g., smartphone GUI) based on the user's interaction behaviors on the new GUI (e.g., smartwatch GUI).

AutoTask "accomplishes unknown tasks in an unknown environment", and all tasks that align with this pattern can benefit from the "explore-learn" strategy outlined in this paper with little effort required from developers or end users. For example, crafting prompts that yield high-performance results can be challenging for end users \cite{dang2022prompt, jiang2022promptmaker, deng2022rlprompt, liu2023wants, wang2023reprompt}. In this problem, the environment is the LLM that has not been fully explored and the task is to generate an effective prompt. However, the effectiveness is not well-defined. AutoTask can attempt an initial prompt to determine the capability boundaries of the LLM and then refine it (similar to the backtracking process described in this paper) based on the responses while accumulating knowledge to enhance its prompt-generation capabilities. Similar concepts \cite{li2023zeroshot} are found in ReAct \cite{yao2022react} and Reflexion \cite{shinn2023reflexion}. However, these approaches do not explicitly summarize knowledge to enhance their capabilities. AutoTask can also extend beyond the digital world into the physical world and be applied in various fields, such as embodied intelligence \cite{brohan2023can, driess2023palm, murali2023improving}.

\section{Limitation \& Future Work}
AutoTask does not interact with users during its execution. However, user commands may be incomplete or ambiguous \cite{pan2023human}, and AutoTask should request clarification or additional information when necessary. Additionally, it can proactively ask questions to prune its GUI exploration based on the user's answers. For example, when a user needs to enable App pinning (Figure \ref{fig: gui-understand}, Page 3), AutoTask may not be familiar with this feature and attempt various incorrect operations. AutoTask can significantly expedite command execution if it proactively asks the user questions like, "Is this feature related to Security?" to gain relevant insights.

AutoTask utilizes an online LLM (gpt-4) service provided by OpenAI via HTTPS requests, and the computational process is slow. Although optimizing the efficiency of LLMs is beyond the scope of this paper, future work could construct smaller and faster models, for example, through techniques such as knowledge distillation \cite{gou2021knowledge}, to reduce waiting time for end users.

Users often have complex intents that cannot be covered by a single GUI task \cite{grudin2019chatbots, luger2016like, fast2018iris}. For example, in the command "Send my schedule to Alice", a voice interface is expected to accomplish two tasks (retrieve the schedule and send a message) sequentially. In future work, AutoTask can be combined with complex task decomposition systems \cite{teevan2016productivity, yang2023auto} to satisfy users' complex needs.

\section{Conclusion}
In this paper, we present AutoTask, a voice command interface that automates voice commands by simulating GUI interactions. To make it applicable across different applications and GUI tasks, AutoTask requires no configuration or modification from developers or end users. Instead, AutoTask explores the GUI to attempt different operation sequences and accumulates knowledge from these explorations to enhance its capabilities. The evaluation study proves the feasibility of this approach: AutoTask performs significantly better than the baseline when no knowledge is accumulated, and knowledge accumulation further improves its performance. AutoTask addresses the special case in the voice assistant domain of accomplishing unknown tasks in an unknown environment, a problem pattern that applies to many other similar scenarios. We hope that AutoTask can inspire future work to apply general artificial intelligence to everyday tasks with little effort from the developers or the end users.

\bibliographystyle{ACM-Reference-Format}
\bibliography{sample-base}

\end{document}